\documentclass[11pt,twoside]{article}
\usepackage{asp2014}

\aspSuppressVolSlug
\resetcounters

\begin{document}

\title{rta-dq-lib: a software library to perform online data quality analysis of scientific data}

\author{Leonardo~Baroncelli$^{1,2}$, Andrea~Bulgarelli$^{1}$, Nicolo'~Parmiggiani$^{1}$, Valentina~Fioretti$^{1}$, Antonio~Addis$^{1}$, Giovanni~De Cesare$^{1}$, Ambra~Di Piano$^{1}$, Vito~Conforti$^{1}$, Fulvio~Gianotti$^{1}$, Federico~Russo$^{1}$, Gilles~Maurin$^{3}$, Thomas~Vuillaume$^{3}$, Pierre~Aubert$^{3}$, Emilio~Garcia$^{3}$, and  Antonio~Zoccoli$^{4}$}

\affil{$^1$National Institute of Astrophysics OAS Bologna, Via P. Gobetti 93/3, 40129, Bologna, Italy; \email{leonardo.baroncelli@inaf.it} }
\affil{$^2$Univ. of Bologna, Bologna, Italy}
\affil{$^3$Laboratoire d'Annecy de Physique des Particules, Univ. Grenoble Alpes, Univ. Savoie Mont Blanc, CNRS, LAPP, 74000 Annecy, France}
\affil{$^4$National Institute of Nuclear Physics, Section of Bologna, Viale Berti Pichat 6/2, 40127 Bologna, Italy}

\paperauthor{Andrea~Bulgarelli}{andrea.bulgarelli@inaf.it}{0000-0001-6347-0649}{INAF}{OAS}{Bologna}{BO}{40129}{Italy}
\paperauthor{Nicolo'~Parmiggiani}{@inaf.it}{}{INAF}{OAS}{Bologna}{BO}{40129}{Italy}
\paperauthor{Valentina~Fioretti}{valentina.fioretti@inaf.it}{0000-0002-6082-5384}{INAF}{OAS}{Bologna}{BO}{40129}{Italy}
\paperauthor{Antonio~Addis}{antonio.addis@inaf.it}{0000-0002-0886-8045}{INAF}{OAS}{Bologna}{BO}{40129}{Italy}
\paperauthor{Giovanni~De Cesare}{@inaf.it}{}{INAF}{OAS}{Bologna}{BO}{40129}{Italy}
\paperauthor{Ambra~Di Piano}{@inaf.it}{}{INAF}{OAS}{Bologna}{BO}{40129}{Italy}
\paperauthor{Vito~Conforti}{@inaf.it}{}{INAF}{OAS}{Bologna}{BO}{40129}{Italy}
\paperauthor{Fulvio~Giannotti}{@inaf.it}{}{INAF}{OAS}{Bologna}{BO}{40129}{Italy}
\paperauthor{Federico~Russo}{@inaf.it}{}{INAF}{OAS}{Bologna}{BO}{40129}{Italy}
\paperauthor{G~Maurin}{}{}{LAPP}{}{}{}{74000}{France}
\paperauthor{T~Vuillaume}{}{}{LAPP}{}{}{}{74000}{France}
\paperauthor{P~Aubert}{}{}{LAPP}{}{}{}{74000}{France}
\paperauthor{Emilio~Garcia}{}{}{LAPP}{}{}{}{74000}{France}
\paperauthor{Antonio~Zoccoli}{zoccoli@bo.infn.it}{}{INFN}{}{}{}{74000}{France}


\begin{abstract}

The Cherenkov Telescope Array (CTA) is an initiative that is currently building the largest gamma-ray ground Observatory that ever existed. A Science Alert Generation (SAG) system, part of the Array Control and Data Acquisition (ACADA) system of the CTA Observatory, analyses online the telescope data - arriving at an event rate of tens of kHz - to detect transient gamma-ray events. The SAG system also performs an online data quality analysis to assess the instruments' health during the data acquisition: this analysis is crucial to confirm good detections. A Python and a C++ software library to perform the online data quality analysis of CTA data, called rta-dq-lib, has been proposed for CTA. The Python version is dedicated to the rapid prototyping of data quality use cases. The C++ version is optimized for maximum performance. The library allows the user to define, through XML configuration files, the format of the input data and, for each data field, which quality checks must be performed and which types of aggregations and transformations must be applied. It internally translates the XML configuration into a direct acyclic computational graph that encodes the dependencies of the computational tasks to be performed. This model allows the library to easily take advantage of parallelization at the thread level and the overall flexibility allow us to develop generic data quality analysis pipelines that could also be reused in other applications.
  
\end{abstract}

\section{Introduction}

In the context of multi-messenger astronomy, an astrophysical source is observed by different kinds of instruments, each one capable of interpreting a different kind of signal (gravitational waves, electromagnetic radiation, neutrinos). For transient sources, this coordinated observation is enabled by science alerts generation systems: they perform an online analysis of the instruments' raw data and they issue a science alert to other astrophysical observatories as soon as they perform a detection. These systems also perform an online data quality analysis in order to assess the instruments' health during the data acquisition: this analysis is crucial to confirm a good detections. We present an online data quality system that has been proposed for the Science Alert Generation system \citep{bul} of the ACADA (Array Control and Data Acquisition) work package of the CTA Observatory. The Cherenkov Telescope Array (CTA) is an initiative that is currently building the largest gamma-ray ground Observatory that ever existed \citep{cta}. It will improve the flux sensitivity by an order of magnitude in the energy range of 20 GeV to 300 TeV, compared to present observatories. The large effective area and field of view coupled with the fast pointing capability make CTA a crucial instrument for transient sources detection and understanding of the physics of short-timescale variability phenomena \citep{fio}. The CTA Observatory will be capable of receiving alerts from external observatories and issuing alerts during observations: the Science Alert Generation (SAG) system, part of the Array Control and Data Acquisition (ACADA) system, analyses online the telescope data - arriving at an event rate of tens of kHz - to detect candidate transients. The SAG system also performs an online data quality analysis: a software library called \textit{rta-dq-lib} has been proposed to manage this computational task.

\section{The data quality analysis requirements}
In this section we describe the requirements that must be fulfilled by the data quality analysis system. Data quality indicators are parameters useful for the purpose of the performance and data quality monitoring. One of the goals of the online data quality analysis is to detect if data quality indicators are within or outside some condition ranges. There are three different condition levels: \textit{optimal}, \textit{warning} and \textit{alarm}. If a data quality indicator falls within the \textit{warning} or \textit{alarm} condition ranges, the Operator and Support Astronomer located in the control room must be notified. If a data quality indicator falls within the \textit{alarm} condition range, the current observation must be interrupted. A data quality indicator can be used to monitor the data coming from auxiliary devices or it can be used to monitor the scientific data coming from the instruments. Another responsibility of the data quality analysis system is to aggregate quality indicator data over time, performing general computations such as summations, averages or root mean squares. Also, it must be capable to compute histograms, correlations and time-vs-value plots. These operations must be performed at the maximum speed. Since there are several data types to process, each one with its own set of data fields, the data quality analysis system must be flexible enough to process every type of input and to implement every possible use case.

\section{The proposed software library}
A Python and a C++ software library to perform online data quality analysis, called rta-dq-lib, has been proposed for CTA. We're currently supporting two versions of the library: a Python version, dedicated to the rapid prototyping of data quality use cases and a C++ version, optimized for maximum performance. This software library is used to perform the monitoring of data quality indicators and to aggregate data in order to compute data transformations (such as distributions and correlations). These operations can be chained together: a data quality indicator can be monitored then aggregated (for several data elements) and its distribution can become a data quality indicator itself that can be monitored in turn. The input source of the library are hdf5 files but we have planned to create a wrapper to be able to read data from an online stream. The data quality use cases can be implemented through XML configuration files that define the operations to be performed: in particular, the library allows the user to define the format of the input data and, for each data field, which data quality indicators must be monitored and which types of aggregations and transformations must be applied. Internally, it translates the XML configuration in a computational direct acyclic graph (check the example in Figure 1) composed by several types of nodes: the \textit{DataSource} nodes acquire the data from hdf5 files, \textit{Executor} nodes aggregate or transform a data field, \textit{QualityCheck} nodes monitor a data quality indicator and \textit{Output} nodes write the results on disk. The \textit{Executor} node can be stateless (without memory) or stateful. The DAG model encodes the dependencies of the computational tasks to be performed and it allows the library to easily take advantage of parallelization at the thread level.

\articlefigure[width=.8\textwidth]{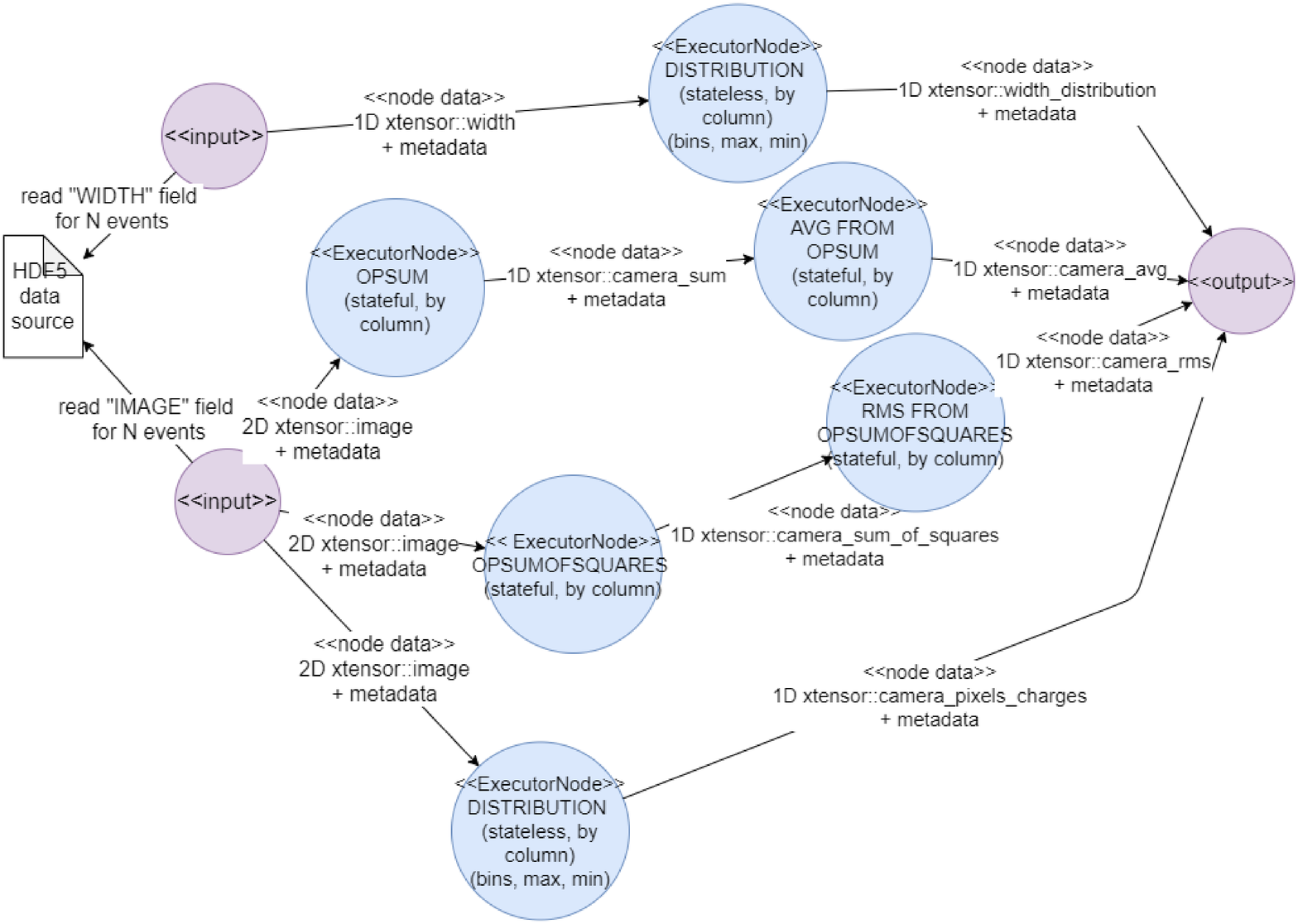}{ex_fig1_reduced}{The representation of the DAG for a given xml configuration. The HDF5 file is composed by several rows and each row represent one data element (or event). Each row (event) is described by several columns (or fields). In this particular example the library reads from the input files two data quality indicators: the "width" and "image" fields. The first represents an Hillas parameter (scalar value) while the second represents the values of the telescope's camera pixels (array value). The "width" data quality indicator is read as an array (one value of "width" field for N events), while the "image" data quality indicator is read as a matrix (1855 values of camera pixels for N events). The library will generate several outputs: an histogram for the distribution of the "width" data quality indicator, a camera plot wich contains the average of the camera pixels, a camera plot which contains the root mean squares of the camera pixels and 1855 histograms representing the distribution of pixels charge for N events.}

\section{Conclusions}
A Python and a C++ software library to perform online data quality analysis, called rta-dq-lib, has been proposed for CTA. It can manage different input data types and provides a configuration mechanism to allow researchers to implements their own data quality analysis use cases writing xml configuration files. It uses a multi-threading approach to speed up the computations in order to be used online in a science alert generation pipeline. The overall flexibility of the library allows the development of generic data quality pipelines that could also be reused in other applications.

\section{Acknowledgment}
This work was conducted in the context of the CTA ACADA Project. \\
We gratefully acknowledge financial support from the agencies and organizations listed here: \begin{verbatim}
    http://www.cta-observatory.org/consortium_acknowledgments
\end{verbatim}

\bibliography{P8-118}

\end{document}